\documentclass[a4paper, 10pt, conference]{ieeeconf}      
\IEEEoverridecommandlockouts	

\usepackage{multirow}
\usepackage{lipsum}
\usepackage{amsmath}
\usepackage{amssymb}
\usepackage{subcaption}
\usepackage{graphicx}
\usepackage{diagbox}
\graphicspath{{./Figures/}}
\usepackage{glossaries}
\usepackage[table,xcdraw]{xcolor}
\usepackage[ruled,vlined]{algorithm2e}
\usepackage[final]{listings}
\usepackage{longtable}
\usepackage[hidelinks]{hyperref}
\usepackage{tikz,pgfplots}

\newacronym{ITS}{ITS}{Intelligent Transportation Systems}
\newacronym{ROS}{ROS}{Robot Operating System}
\newacronym{URDF}{URDF}{Unified Robot Description Format}
\newacronym{SLAM}{SLAM}{Simultenous Localization and Mapping}
\newacronym{JSON}{JSON}{JavaScript Object Notation}
\newacronym{LDS}{LDS}{Laser Distance Sensor}
\newacronym{iCab}{iCab}{Intelligent Campus Automobile}
\newacronym{AV}{AV}{Autonomous Vehicle}
\newacronym{HMI}{HMI}{Human Machine Interface}
\newacronym{PAFs}{PAFs}{Part Affinity Fields}
\newacronym{VRUs}{VRUs}{Vulnerable Road Users}

\title{\LARGE \bf Vehicle Automation Field Test: Impact on Driver Behavior and Trust}

\author{Walter Morales-Alvarez$^{1}$ \emph{Student Member, IEEE}, Nikita Smirnov$^{1, 2}$, Elmar Matthes$^{3}$  \\
Cristina Olaverri-Monreal$^{1}$ \emph{Senior Member, IEEE}%
\thanks{$^1$ Johannes Kepler University Linz, Austria; Chair  Sustainable Transport Logistics 4.0. \texttt{\{walter.morales\_alvarez, nikita.smirnov, cristina.olaverri-monreal\}@jku.at}}%
\thanks{$^2$ Ural Federal University, Department of Communications Technology.}
\thanks{$^3$ IAV GmbH, Carnotstrasse 1, 10587 Berlin, Germany \texttt{elmar.matthes@iav.at}}%
}

\newcommand\copyrighttext{%
  \footnotesize \textcopyright 2020 IEEE. Personal use of this material is permitted. Permission from IEEE must be obtained for all other uses, in any current or future   media, including reprinting/republishing this material for advertising or promotional purposes, creating new collective works, for resale or redistribution to servers or lists, or reuse of any copyrighted component of this work in other works.  DOI: \href{https://ieeexplore.ieee.org/document/9304624}{10.1109/IV47402.2020.9304624} }

\newcommand\copyrightnotice{%
\begin{tikzpicture}[remember picture,overlay]
\node[anchor=south,yshift=10pt] at (current page.south) {\fbox{\parbox{\dimexpr\textwidth-\fboxsep-\fboxrule\relax}{\copyrighttext}}};
\end{tikzpicture}%
}

\begin{document}

\maketitle
\copyrightnotice
\thispagestyle{empty}
\pagestyle{empty}

\hyphenation{obtained}
\hyphenation{automated}
\hyphenation{Climate}
\hyphenation{Sustainable}
\hyphenation{Innovation}
\hyphenation{Ministry}

\begin{abstract}

With the growing technological advances in autonomous driving, the transport industry and research community seek to determine the impact that autonomous vehicles (AV) will have on consumers, as well as identify the different factors that will influence their use. Most of the research performed so far relies on laboratory-controlled conditions using  driving simulators, as they offer a safe environment for testing advanced driving assistance systems (ADAS). 
In this study we analyze the behavior of drivers that are placed in control of an automated vehicle in a real life driving environment. The vehicle is equipped with advanced autonomy, making driver control of the vehicle unnecessary in many scenarios, although a driver take over is possible and sometimes required. In doing so, we aim to determine the impact of such a system on the driver and their driving performance.  To this end road users' behavior from naturalistic driving data is analyzed focusing on awareness and diagnosis of the road situation. Results showed that the road features determined the level of visual attention and trust in the automation. They also showed that the activities performed during the automation affected the reaction time to take over the control of the vehicle.
\end{abstract}


\section{Introduction}
\label{sec:introduction}

There has been a drastic increase in the number of vehicular systems on the market that rely on some degree of automation. These advanced driving assistance systems (ADAS) enhance driver perception and augment the driver’s awareness of the surrounding environment, thereby helping to increase road safety ~\cite{michaeler20173d}. However, they are not necessarily accepted and utilized by all users, reflecting the perceived trustworthiness of the automation and the skepticism of drivers. This is why user validation of human-centric technology is a pressing issue that is of particular interest in current research.

In high automation, the automated driving system maintains control of the vehicle~\cite{allamehzadeh2016automatic} and the driver is not required to monitor the road~\cite{gasser2012bast}. As a consequence, drivers are free to engage in non-driving tasks~\cite{merat2012highly}. 
Current prototypes are able to control braking, acceleration and steering but drivers still need to monitor the road, as a response from the driver to a Take Over Request (TOR) is expected. 

Today, phones are multi-functional, ubiquitous devices that are capable of connect with other devices in a variety of ways. Modern phones have applications that generate the need for people to check them periodically; so much so that studies show that a large part of drivers use phones while they are driving \cite{White2010}, \cite{Gras2007}. This practice is extremely dangerous in traditional vehicles. However, with the incursion of highly automated vehicles, drivers will be able to safely deviate their attention from the road.

Therefore, the advent of autonomous driving represents an opportunity for increased road safety, as the automation will make superfluous driver intervention in the control of the vehicle~\cite{olaverri2016autonomous}. 
However, the majority of autonomous vehicles currently in operation are used to transport goods or persons within the boundaries of specific industrial sites. In this kind of environment human-machine interaction occurs under relatively controlled conditions and the operators are familiar with the vehicles and the way they function.

In a real road situation the automation contributes to a decrease in driving workload and a consequent reduction in driver situational awareness, which needs to be taken into account when a vehicle control is expected from the driver. Therefore, in high automation driving, Driver State Monitoring Systems (DSMS) play a crucial role in determining whether or not the driver is prepared to take control of the vehicle and to respond to the TOR in the most appropriate manner~\cite{8500367, 8317925}. An example of DSMS  that is employed in some vehicles is a signal that is automatically activated if the driver does not have their hands on the steering wheel for a certain period of time. 

In this work we determine the risks of using mobile phones in a field test under real driving conditions while the vehicles' lateral control is managed by the automation. We examine the impact of the automation on driver behavior and trust by analyzing gaze direction and frequency, as well as the reaction time to a TOR, and formulate the following research question: 

Do drivers trust the automated capabilities of the vehicles when they are performing secondary tasks? 

by defining the following hypotheses:

\begin{enumerate}
\item Road features (i.e road curves) and secondary tasks do not affect trust in  the automation.
\item The reaction time to take back control of the vehicle does not depend on the kind of secondary task the driver is performing during the automation.
\end{enumerate}

The remainder of the paper is organized as follows: the next section describes related work in the field; section~\ref{sec:fieldtest} details the  field  test.  Section~\ref{sec:analysis} presents  the  method used to  assess  the  data collected;  section~\ref{sec:results} presents  the  obtained  results;  and finally, section~\ref{sec:conclusion} discusses and concludes the work.

\section{Related Work}
\label{sec:relatedwork}


A large number of studies have been performed related to the monitoring and assessment of drivers in different situations with the goal of developing systems that optimize autonomous vehicles. An example is the study in \cite{allamehzadeh2017cost} that focused on creating a system that detected the state of drowsiness of the driver, as well as pedestrian and surrounding vehicles, by using the data acquired through the cameras of a smart phone. Based on this information the system could issue a TOR depending on the road situation and the current state of the driver. 
In the same line of research, the authors in \cite{Bylykbashi2020} acquired data from monitoring the driver and the environment to determine the driver's state of situational awareness using fuzzy logic. In a further work \cite{Goncalves2015}, the architecture of a system that calculated the driver's state in a highly automated vehicle was presented. The system determined whether it was safe to give the driver control of the vehicle or if the vehicle itself should take care of the corresponding maneuver. 

In most studies different external factors were manipulated to determine how they affected the state of the driver and their driving performance. For example, in \cite{Takada2019} and \cite{Noh2019} a variety of images or messages were conveyed to the driver and their cognitive load and stress level were measured afterward using ECG and ECC signals, respectively.
 
Warnings regarding road safety such as obstacles on the street or a failure in the automation system resulted in an increase of trust \cite{Petersen2019,Beller2013}.  

Also related to trust, the works in \cite{Verberne2012} and \cite{Xiong2012} studied the effects of an adaptive cruise control system on a sample of participants, showing the results that overtrust affected driving performance negatively.

Furthermore \cite{Xiong2012} analyzed trust in the automation using simulated vehicles in platoon mode. Results showed that the level of trust depended directly on the efficiency of the vehicle automation.

The existing studies based their data on driver behavior in real or virtual situations using conventional or virtual vehicles. Studies on driver behavior in real road situations that concern automated vehicles are very scarce due to the complexity and the risk factors that they involve. To contribute to the field of research we perform in this study a field test under driving conditions in which drivers' lateral control is managed by the automation.

\section{Field Test Description}
\label{sec:fieldtest}

\subsection{Participants}

The sample of participants consisted of 15 persons with valid German driving licenses and varying experience with lane keeping systems (LKS). Of the 15 volunteers two participants were discarded because of the noise of their recorded data. 

The remaining participants consisted of 5 females and 8 males with an average age of 28.26 years (SD = 9.05) and an average driving experience of 10.15 years (SD = 9.05). Table~\ref{table:paricipants} provides a visualized summary of the sample characteristics.

\begin{table}
	\centering
	\scriptsize
	\caption{Participant distribution regarding sex, age and driving experience}
	\label{table:paricipants}
		\begin{tabular}{|p{8cm}|}
		\hline  
		\textbf{Sample (8 males, 5 females)}\\
	\end{tabular}
	\begin{tabular}{|p{2.93cm}|p{2.1cm}|p{2.1cm}|}
		\hline
		 & Mean & SD \\
		\hline
		Age & 28.26 & 9.05 \\
		\hline
		Driving Experience & 10.15 &9.05\\
	\end{tabular}
	\begin{tabular}{|p{8cm}|}
		\hline
		\textbf{LKS Knowledge}\\
	\end{tabular}
	\begin{tabular}{|p{1.255cm}|p{1.25cm}|p{1.25cm}|p{1.25cm}|p{1.25cm}|}
		\hline
		None&Few&Neutral&Much&Confident\\
		\hline
		40\%&13.3\%&26.7\%&13.3\%&6.7\%\\
		\hline
	\end{tabular}
\end{table}

\subsection{Test Procedure}

The participants had to drive for a certain time period with an automated lane keeping system activated while performing several predefined secondary tasks. They were instructed to accelerate until they reached 60 km/h, at which time the vehicle's lane keeping system was activated and the secondary tasks started. 
The tasks to be performed forced them to deviate their eyes from the road and to remove their hands from the steering wheel. 

Prior to the test, they were introduced to the experimental procedure and instructed to keep the system active until a TOR was triggered. This take over request occurred 20 seconds after the driver released their hands from the steering wheel. 

A training session was performed before the test so that the participants could get accustomed to the vehicle and its system. During the driving experiment each participant completed three different tasks on the same test track during a total time of 10 minutes. 

At the end, each participant completed a post-task questionnaire to determine their situational awareness and self-perceived trust in the system.


\subsection{Secondary Tasks}

To assess the impact of the driver assistance system in a real life situation, the secondary tasks were designed to replicate everyday actions that usually occur on the road and are as follows:

\begin{itemize}
	\item \textbf{Baseline:} LKS activated without any task. 
	\item \textbf{Visual:} Reading a text (aloud) from a smart phone. 
	\item \textbf{Visual and Manual:} Writing a given text on a smart phone.
\end{itemize}   

\begin{figure}
	\centering
	\includegraphics[width=0.48\textwidth]{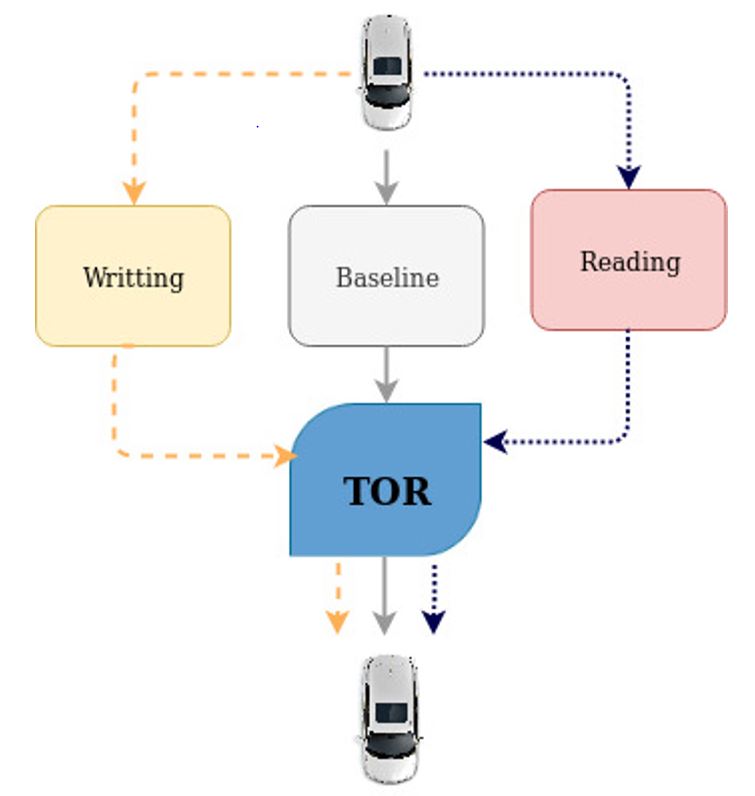}
	\caption{ Illustration of the driving experiment procedure. }
	\label{fig:test}
\end{figure}

\begin{figure}
	\centering
	\includegraphics[width=0.48\textwidth]{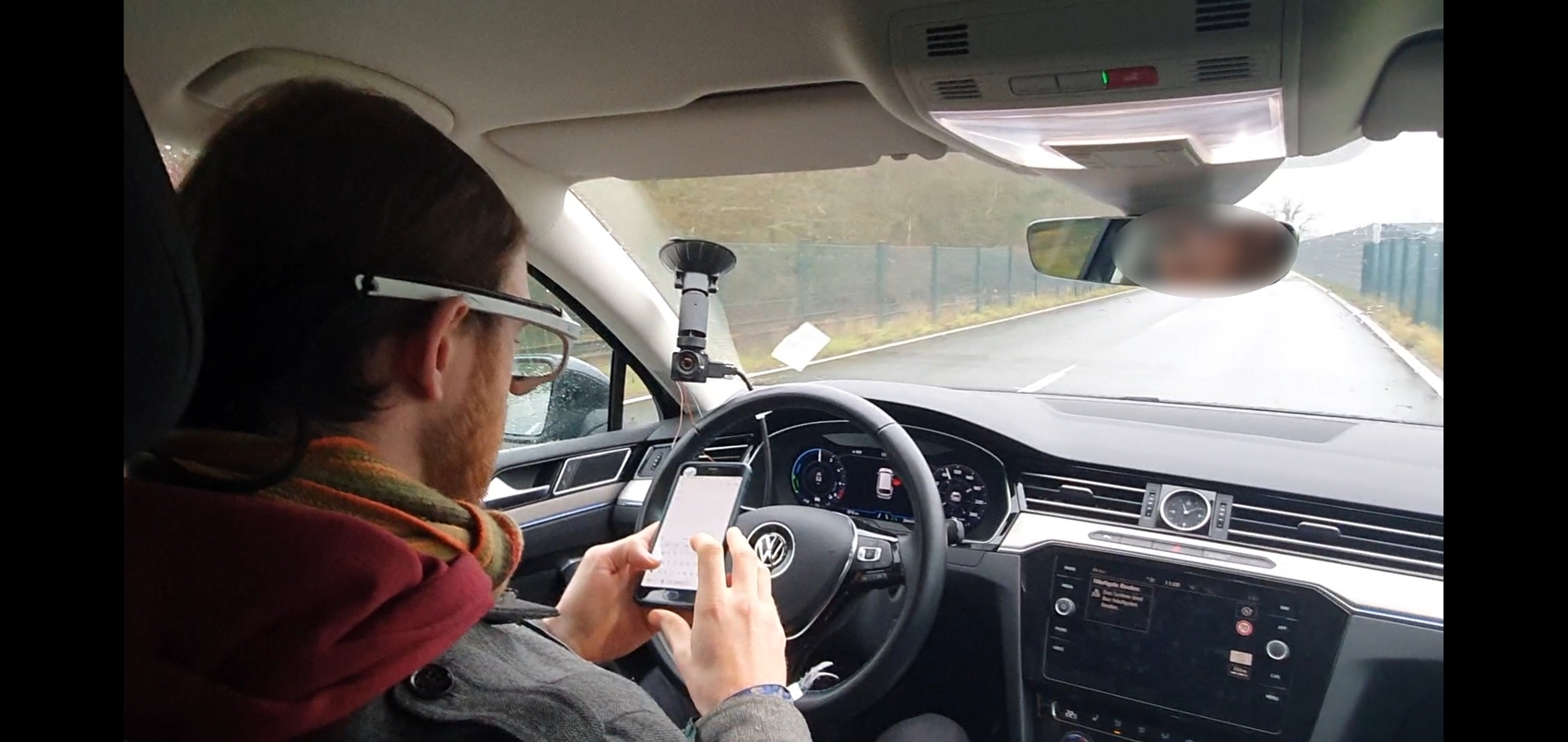}
	\caption{ Participant performing one of the secondary tasks while the vehicle was controlled by the lane keeping system. }
	\label{fig:participant}
\end{figure}

Both texts were tongue twisters in the native language of the participants (German) in order to slightly increase the mental workload of the task. Figure~\ref{fig:test} depicts the test procedure.
Figure~\ref{fig:participant} shows a participant performing one of the secondary tasks while the vehicle was controlled by the lane keeping system.

\subsection{Apparatus and Test Scenario}

\begin{figure}
	\centering
	\includegraphics[width=0.48\textwidth]{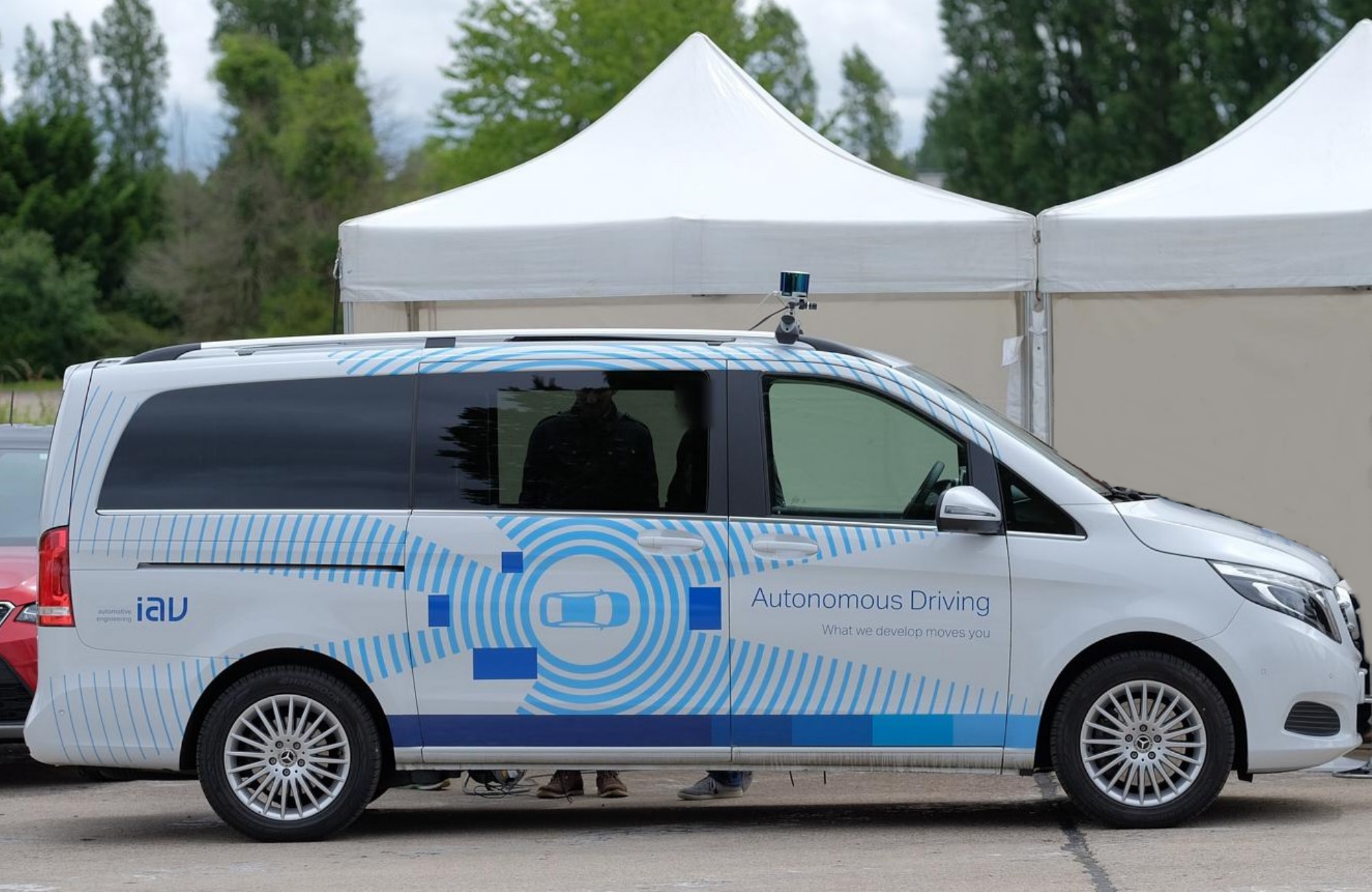}
	\caption{ Vehicle with automated capabilities that was used to perform the tests. }
	\label{fig:apparatus}
\end{figure}

The test were conducted in a test track using a research vehicle of the Automotive Technology Manufacturers IAV GmbH (Figure~\ref{fig:apparatus}) that was equipped with a a LKS and other equipment to sense the environment (e.g camera). An internal camera was additionally available to monitor the head pose of the participants. 
The participants carried out the secondary task on a Samsung S10 plus, which has a 6.4-inch screen using a QWERTY keyboard, instead of the German QWERTZ keyboard, to increase the mental workload of the task.

The scenario chosen to conduct the tests was located in Germany. This track consists of a closed road with two lanes (one for each direction), with a slight curvature of approximately 850 m. For safety reasons, the track was reserved for these tests, without traffic being involved.

\section{Driver Assessment Procedure}
\label{sec:analysis}

In order to test the hypothesis 1 and the level of trust in the automation, we focused on participant gaze direction and frequency, as well as road features since in order to perform the tasks they needed to deviate the eyes from the road and focus on the smart phone. 
To test the hypothesis 2 we calculated the reaction times necessary to take back control of the vehicle for both types of secondary task performed during the automation.

\subsection{Data Acquisition}
To this end the following dependent variables were acquired through the external modules of the vehicle (CAN and camera):

\begin{itemize}
	\item \textbf{Eyes on road frequency:} Defined as the number of times drivers deviated their look from the phone screen to the road.
	\item \textbf{Eyes on road average time:} The average time that the participants look at the road.
		\item \textbf{Reaction time:} Defined as the time it took the participant to put their hands on the steering wheel from the moment in which a TOR was triggered.
\end{itemize} 

The data from the sensors installed for these tests was obtained through a private framework developed by the company IAV.

\subsection{Data Processing and Analysis}

To process and further analyze the obtained data it was converted to the format of Robotic Operative System 2 (ROS 2). 

The data processing was done semi-automatically by programming two ROS 2 nodes that subscribe to the camera and CAN data independently. The node that was connected to the camera was in charge of showing the recorded images and from these it was observed when the participant deviated their gaze from the road and / or phone. The node was in charge of writing down the times between these two frames, which were pointed out by the observer through the computer keyboard. 

In relation to the CAN data, it contained the messages that indicated when the acoustic signal for the TOR occurred and when the participant had taken control of the vehicle. Therefore, to determine the driver's reaction time, it was only necessary to program a node that would be in charge of measuring the time between these two messages.

\begin{figure*}
	\begin{subfigure}{\textwidth}
		\includegraphics[width=0.33\textwidth]{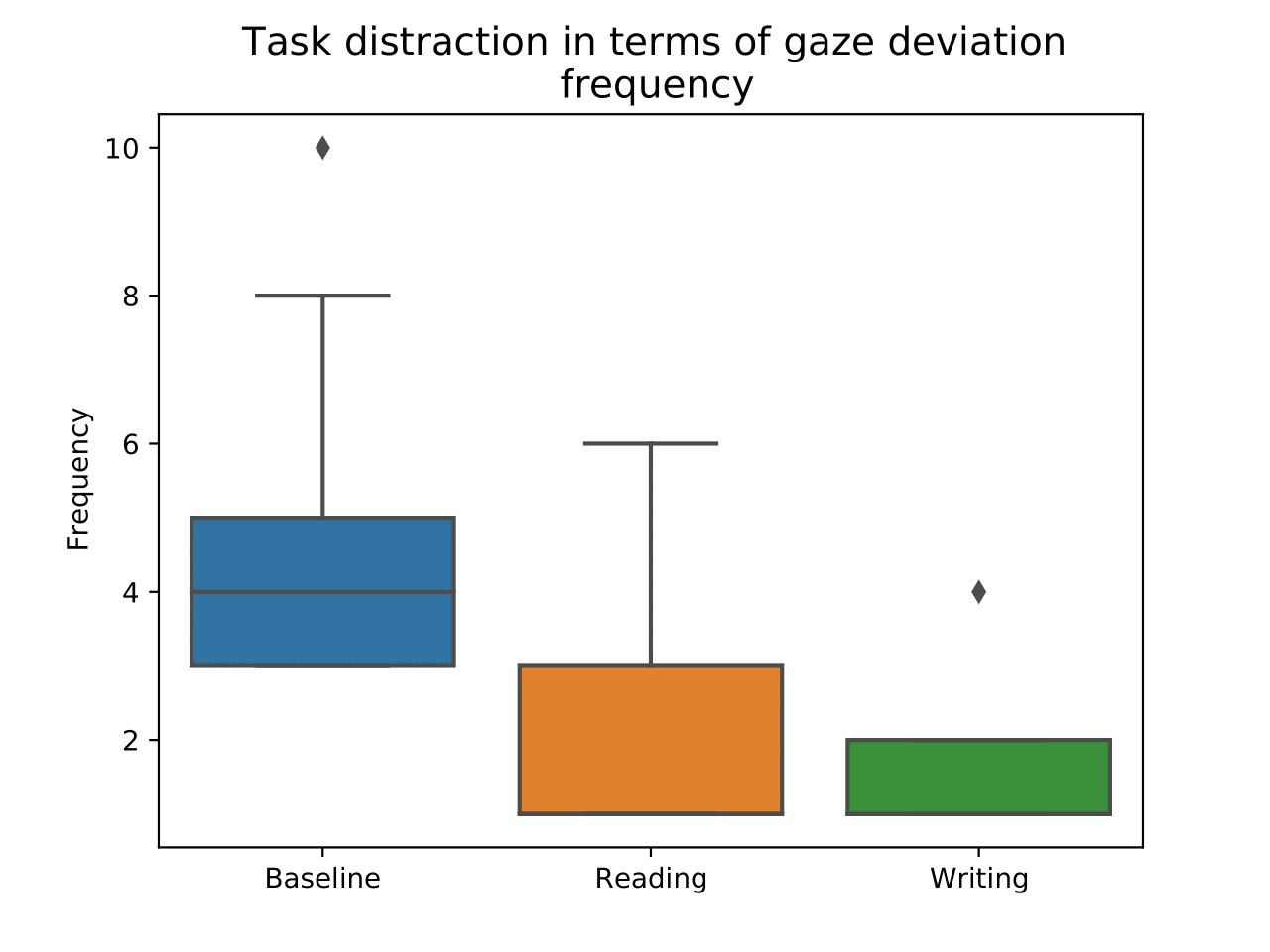}
		\includegraphics[width=0.33\textwidth]{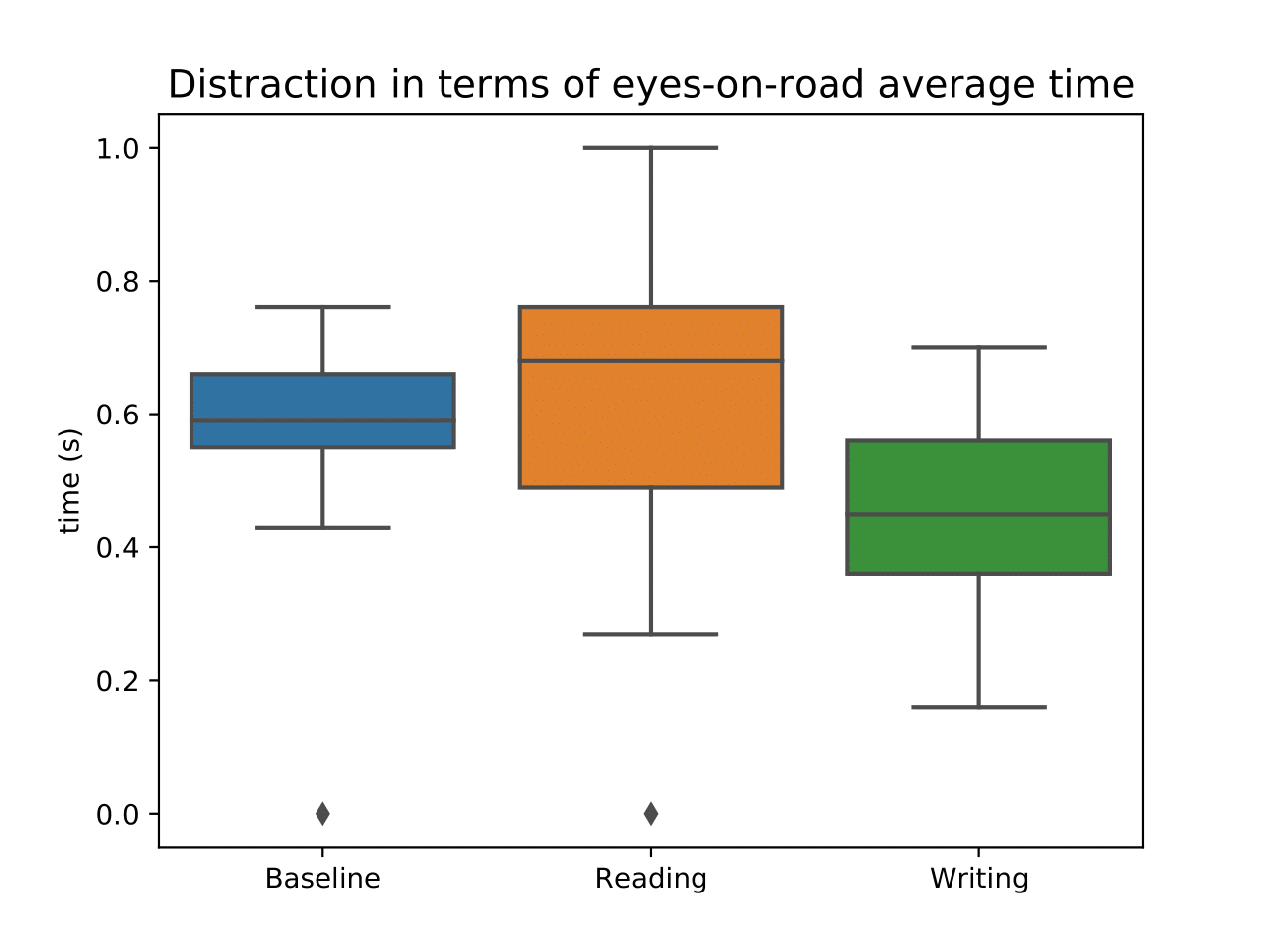}
		\includegraphics[width=0.33\textwidth]{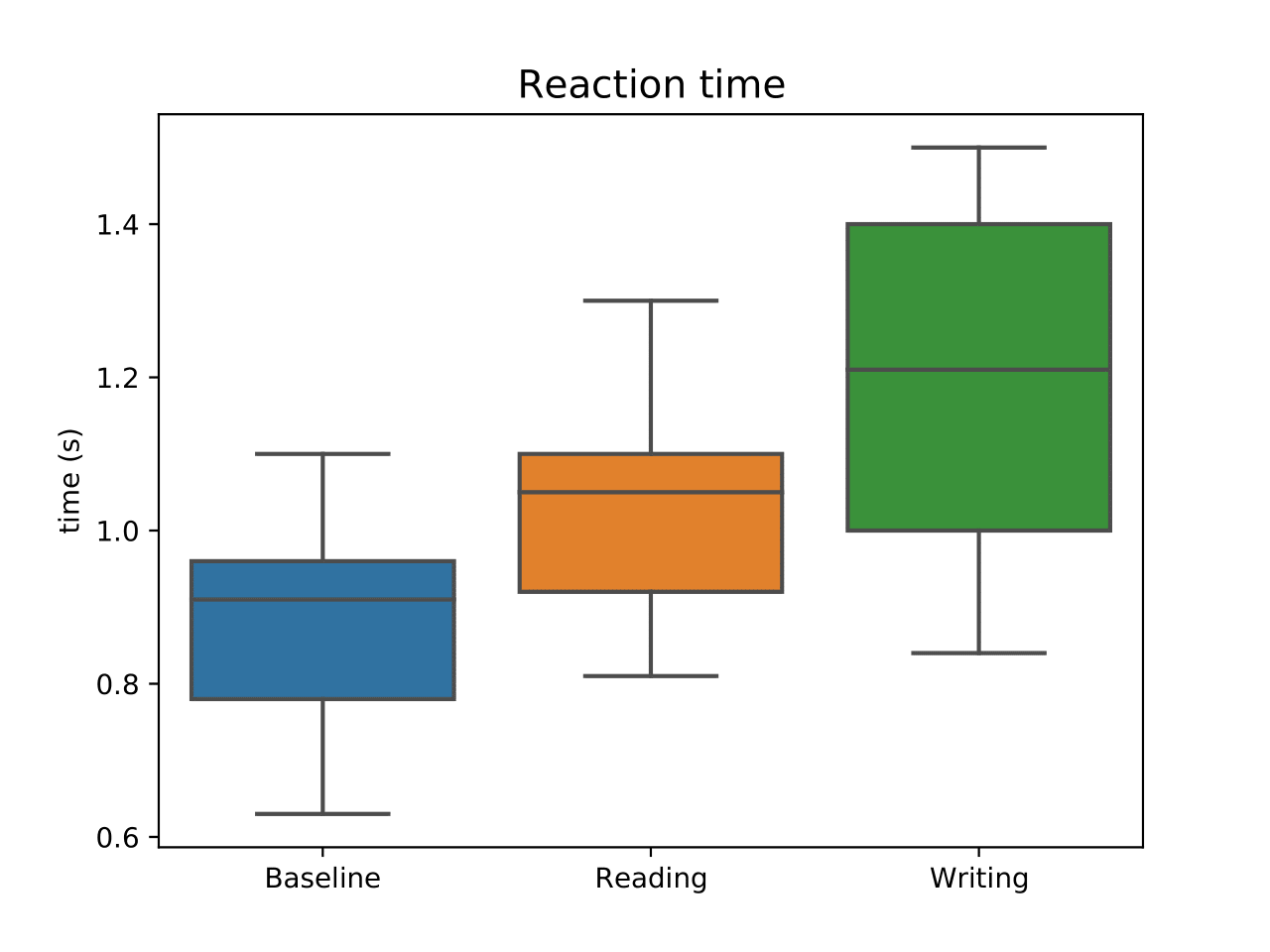}
		\caption{}
		\label{fig:plots:a}
	\end{subfigure}
	\begin{subfigure}{\textwidth}
		\includegraphics[width=0.33\textwidth]{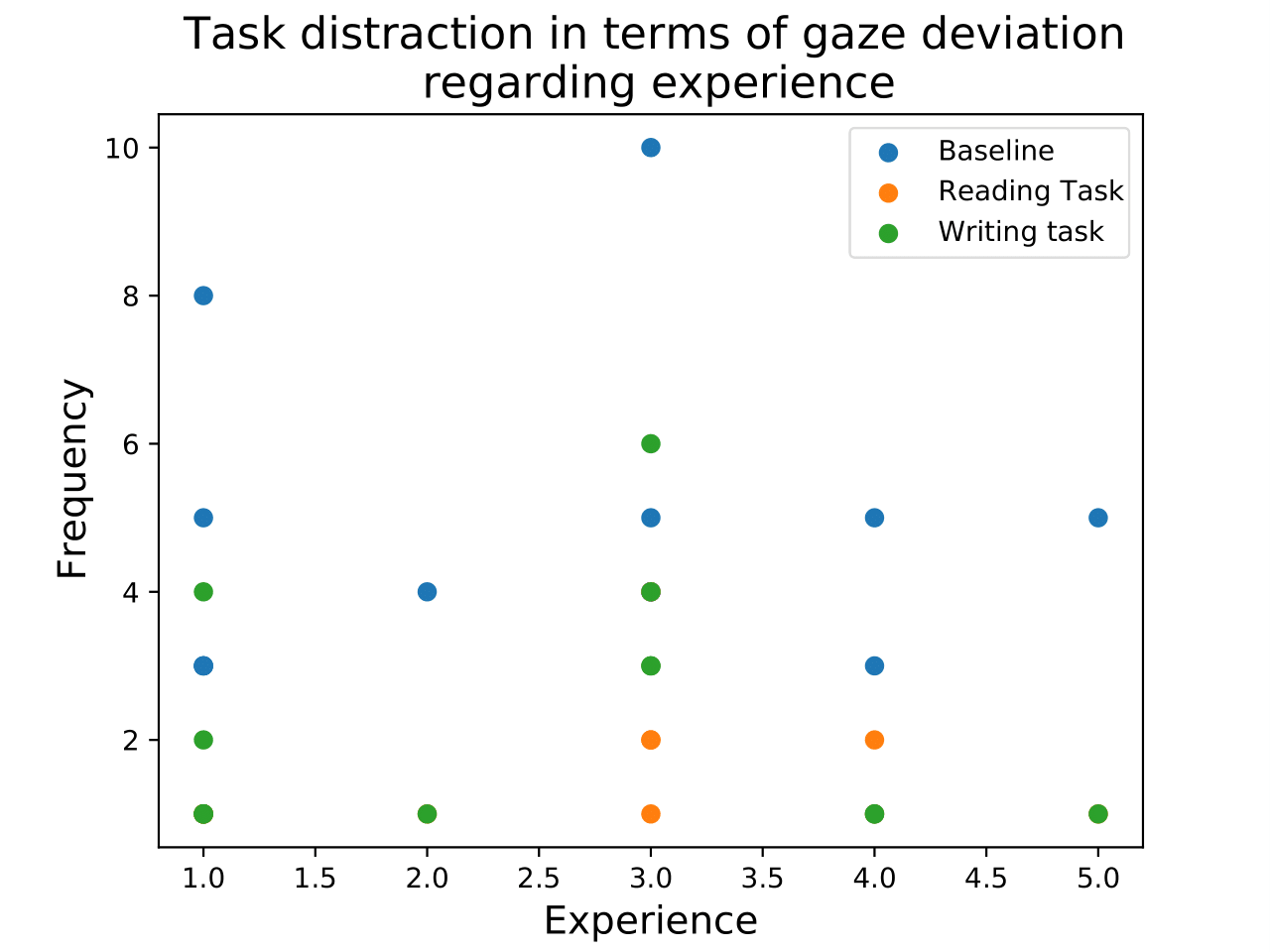}
		\includegraphics[width=0.33\textwidth]{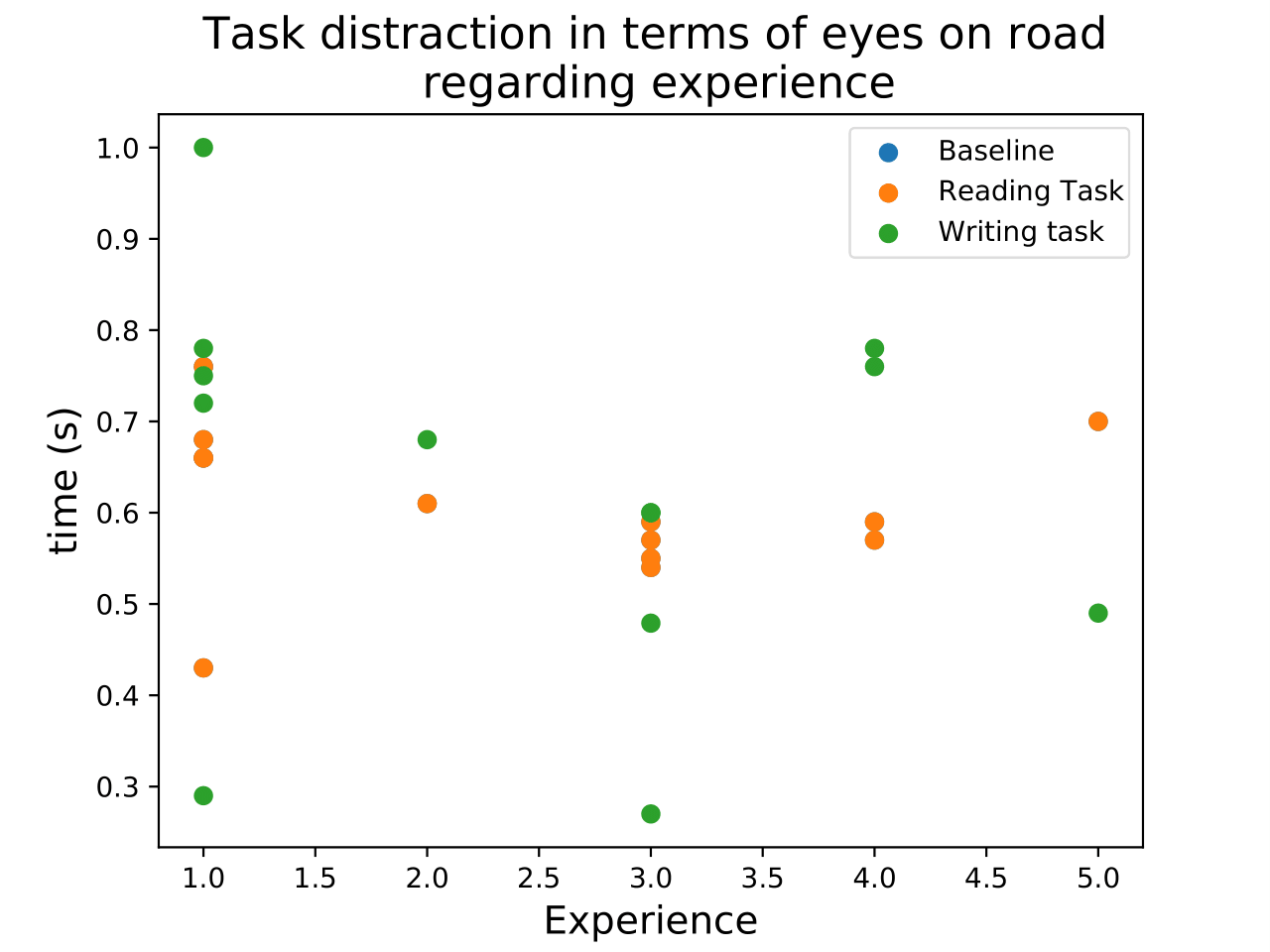}
		\includegraphics[width=0.33\textwidth]{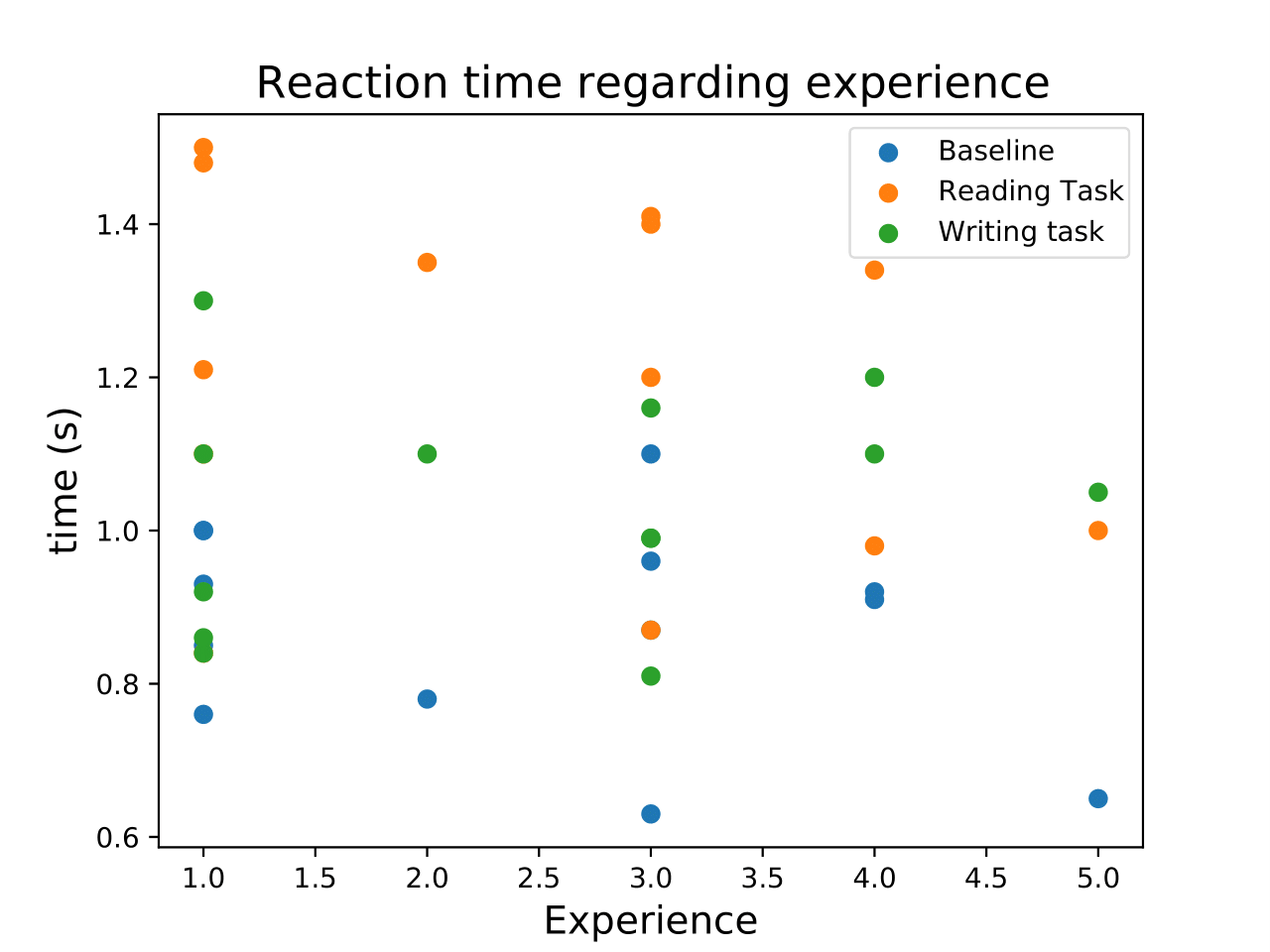}
		\caption{ }
		\label{fig:plots:b}
	\end{subfigure}
	\caption{Results obtained from the data acquired in each test, including: a) results regarding the gaze deviation, eyes-on road and reaction time depending on the secondary task and b) distraction and reaction time depending on driving experience.}
	\label{fig:plots}
\end{figure*}

To analyze the dependent variables mentioned above, we used the SPSS program where the T-student test was done to determine if there is a relationship between the dependent variables and the secondary task performed. 

Furthermore, given that one of the questions asked to the participants was their previous experience with automated capabilities in vehicles, we analyzed a potential correlation between this reported experience and the dependent variables obtained in the experiments.

Finally, we analyzed the relationship between the frequency of diverting the visual attention from the secondary task and features of the road, namely a curve.


Subsequently, we performed a Person Chi Square analysis of the participants post-questionnaire answers in order to obtain a qualitative result of the trust in the systems automated capabilities. The range of the answers varied between 1 (low) and 7 (high).

\section{Results}
\label{sec:results}

The results regarding the drivers’ behavioral patterns are illustrated in Table~\ref{table:t-test}. It can be observed that there is a significant statistical relationship between the frequency with which drivers diverted their gaze and visual attention from the phone screen to the road and whether they were engaged in the secondary tasks or not (baseline). The eyes on road frequency was higher when they were performing the reading tasks than with the writing task.

\begin{table}
	\centering
	\scriptsize
	\caption{Driver's behavior depending on the type of secondary task performed }
	\label{table:t-test}
	\begin{tabular}{|p{1.3cm}|p{1.84cm}|p{1.73cm}|p{1.86cm}|}
		\hline
		Metric&Baseline&Read secondary task&Write secondary task\\
	\end{tabular}
	\begin{tabular}{|p{1.3cm}|p{0.7cm}|p{0.7cm}|p{0.7cm}|p{0.7cm}|p{0.7cm}|p{0.6cm}|}
		\hline
		&Mean&SD&Mean&SD&Mean&SD\\
		\hline
		Eyes on road frequency&4.769&2.087&2.231&1.640&1.461&0.877\\
		\hline
		Eyes on road average time&0.610&0.087&0.641&0.211&0.455&0.154\\
		\hline
		Reaction time&0.738&0.138&1.031&0.148&1.206&0.229\\
		\hline
	\end{tabular}
	\begin{tabular}{|p{8.02cm}|}
		\hline
		T-test ($\alpha$ =0.05)\\
	\end{tabular}
	\begin{tabular}{|p{1.3cm}|p{1.84cm}|p{1.73cm}|p{1.86cm}|}
		\hline
		Metric&Baseline vs read secondary task&Baseline vs write secondary task&Read secondary task vs write secondary task\\
	\end{tabular}
	\begin{tabular}{|p{1.3cm}|p{0.7cm}|p{0.7cm}|p{0.7cm}|p{0.7cm}|p{0.7cm}|p{0.6cm}|}
		\hline
		&t(12)&p&t(12)&p&t(12)&p\\
		\hline
		Eyes on road frequency&4.365&0.001**&7.217&0.000**&2.379&0.035*\\
		\hline
		Eyes on road average time&-0.481&0.640&2.600&0.025*&2.793&0.017*\\
		\hline
		Reaction time&-3.454&0.005**&-5.462&0.001**&-3.461&0.005**\\
		\hline
	\end{tabular}
\end{table}

With regard to the average time that drivers looked at the road and diverted their attention from the phone, there is no statistically significant relationship between the baseline condition and the reading condition. However, there was a statistically significant difference between the baseline condition or the reading condition and the writing condition. Drivers diverted their gaze away from the phone for a shorter period of time when writing on the phone than in the other cases.

Results regarding the reaction time showed that there is a statistical significant relationship between this dependent variable and the task with which the driver was engaged. 

The results showed that in most cases the participants needed longer to take control of the vehicle when they were writing on a phone than in the other cases, and at the same time a slower reaction time was obtained when reading than without performing a secondary task under the baseline condition (See Figure~\ref{fig:plots:a}).

Regarding a correlation between the reported experience with automated systems and the dependent variables obtained in the experiments,  Figure~\ref{fig:plots:b} shows that there was no relationship between the behavior of the drivers in the tests and the experience with the lane keeping system. The correlation coefficients obtained were in any case below 0.13 ($r\leq |0.13|$). 

\begin{figure}
	\centering
	\includegraphics[width=0.45\textwidth]{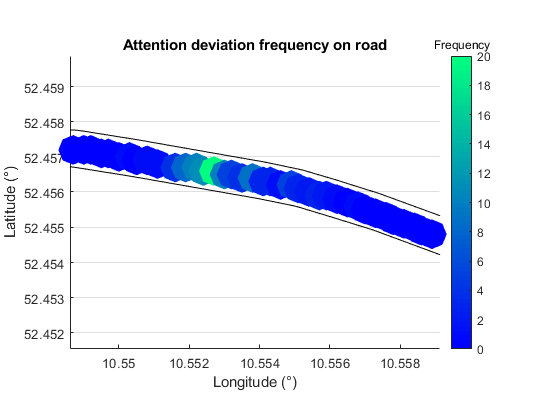}
	\caption{Driver's attention deviation regarding road feature.}
	\label{fig:road}
\end{figure}

The results regarding the relationship between a specific road feature such as curves on the road and the visual attention deviation from the phone in Figure~\ref{fig:road} show an increase in frequency when drivers approached a curve. 

Finally, the level of trust reported by the participants in the experiment can be seen in the Table~\ref{table:chi1}. Results showed a statistically significant relationship between the perceived trustworthiness of the automation and the test performed, as the participants reported higher levels of trust in the automation under baseline conditions and less under writing conditions. 

Based on the results we can reject the defined hypotheses 1 and 2 and affirm that the type of secondary task and road features affect trust in the automation and that the reaction time to take back the control of the vehicle depends on the kind of secondary task performed during the automation.

\begin{table}
	\centering
	\scriptsize
	\caption{Driver's reported trust regarding each secondary task}
	\label{table:chi1}
	\begin{tabular}{|p{1.3cm}|p{1.84cm}|p{1.73cm}|p{1.84cm}|}
		\hline
		&Baseline&Read secondary task&Write secondary task\\
		\hline
		Trust&11&6&3\\
		\hline
		No trust&2&7&10\\
	\end{tabular}
	\begin{tabular}{|p{8.02cm}|}
		\hline
		$\chi{}^2$ test ($\alpha$ =0.05)\\
	\end{tabular}
	\begin{tabular}{|p{2.38cm}|p{2.38cm}|p{2.38cm}|}
		\hline
		Baseline vs Read secondary task&Baseline vs Write secondary task&Read secondary task vs Write secondary task\\
	\end{tabular}
	\begin{tabular}{|p{0.97cm}|p{0.97cm}|p{0.97cm}|p{0.97cm}|p{0.97cm}|p{0.97cm}|}
		\hline
		(1,N=26)&p&(1,N=26)&p&(1,N=26)&p\\
		\hline
		4.923&0.0265&9.904&0.001**&4.181&0.0409*\\
		\hline
	\end{tabular}
\end{table}

\section{Conclusion, Discussion and Future Work}
\label{sec:conclusion}

In this paper we presented a study under real field test conditions of the impact of automated driving systems on driver behavior and trust. 

We measured visual attention and gaze with respect to secondary tasks and road features, as well as the reaction time to a TOR, and concluded that the secondary task performed by the driver during the automation affected driver behavior and reaction time. 

When the drivers were not engaged in a task, their gaze remained on the road. During secondary tasks, the participants’ gaze behavior indicates that they were immersed in their tasks, especially so for the writing task, and therefore trusting of the automation and not worried about a potential failure. It was observed how smart phones make people forget that they are on the road, the task of writing a text causing one to have their eyes off road the longest. This may be because this task requires more concentration to be completed or involves use of one’s hands.
However, this objective results did not correspond with the skepticism that resulted from the analysis of the qualitative data  
that showed that as the task became more complicated, participants reported a lower level of trust in the lane keeping automation.

In more challenging road conditions, the probands care less about the secondary task and trust the system less. Regardless of which secondary task drivers were performing, in most cases they were diverting their attention to the road when the vehicle was approaching a curve, implying a heightened degree of concern about the vehicle system when handling this type of road feature.

Writing on a phone, distracted the drivers from observing their environment, demonstrating how unsafe this task is in traditional vehicles. 

The previous experience of some participants with automated vehicular systems did not affect their behavior. 

Future work will focus on using specialized sensors to monitor drivers, such as eye tracking devices or sensors to measure stress, as well as external data regarding the driving scenario and conditions.

\section*{ACKNOWLEDGMENT}
This work was supported by the Austrian Ministry for Climate Action, Environment, Energy, Mobility, Innovation and Technology (BMK) Endowed Professorship for Sustainable Transport Logistics 4.0.
\bibliographystyle{IEEEtran}
\bibliography{paper}

\begin{thebibliography}{10}
\providecommand{\url}[1]{#1}
\csname url@samestyle\endcsname
\providecommand{\newblock}{\relax}
\providecommand{\bibinfo}[2]{#2}
\providecommand{\BIBentrySTDinterwordspacing}{\spaceskip=0pt\relax}
\providecommand{\BIBentryALTinterwordstretchfactor}{4}
\providecommand{\BIBentryALTinterwordspacing}{\spaceskip=\fontdimen2\font plus
\BIBentryALTinterwordstretchfactor\fontdimen3\font minus
  \fontdimen4\font\relax}
\providecommand{\BIBforeignlanguage}[2]{{%
\expandafter\ifx\csname l@#1\endcsname\relax
\typeout{** WARNING: IEEEtran.bst: No hyphenation pattern has been}%
\typeout{** loaded for the language `#1'. Using the pattern for}%
\typeout{** the default language instead.}%
\else
\language=\csname l@#1\endcsname
\fi
#2}}
\providecommand{\BIBdecl}{\relax}
\BIBdecl

\bibitem{michaeler20173d}
F.~Michaeler and C.~Olaverri-Monreal, ``{3D driving simulator with VANET
  capabilities to assess cooperative systems: 3DSimVanet},'' in
  \emph{Intelligent Vehicles Symposium (IV)}.\hskip 1em plus 0.5em minus
  0.4em\relax IEEE, 2017, pp. 999--1004.

\bibitem{allamehzadeh2016automatic}
A.~Allamehzadeh and C.~Olaverri-Monreal, ``{Automatic and manual driving
  paradigms: Cost-efficient mobile application for the assessment of driver
  inattentiveness and detection of road conditions},'' in \emph{2016 IEEE
  Intelligent Vehicles Symposium (IV)}.\hskip 1em plus 0.5em minus 0.4em\relax
  IEEE, 2016, pp. 26--31.

\bibitem{gasser2012bast}
T.~M. Gasser and D.~Westhoff, ``Bast-study: Definitions of automation and legal
  issues in germany,'' in \emph{Proceedings of the 2012 road vehicle automation
  workshop}.\hskip 1em plus 0.5em minus 0.4em\relax Automation Workshop, 2012.

\bibitem{merat2012highly}
N.~Merat, A.~H. Jamson, F.~C. Lai, and O.~Carsten, ``Highly automated driving,
  secondary task performance, and driver state,'' \emph{Human factors},
  vol.~54, no.~5, pp. 762--771, 2012.

\bibitem{White2010}
K.~M. White, M.~K. Hyde, S.~P. Walsh, and B.~Watson, ``{Mobile phone use while
  driving: An investigation of the beliefs influencing drivers' hands-free and
  hand-held mobile phone use},'' \emph{Transportation Research Part F: Traffic
  Psychology and Behaviour}, vol.~13, no.~1, pp. 9--20, 2010.

\bibitem{Gras2007}
M.~E. Gras, M.~Cunill, M.~J. Sullman, M.~Planes, M.~Aymerich, and
  S.~Font-Mayolas, ``{Mobile phone use while driving in a sample of Spanish
  university workers},'' \emph{Accident Analysis and Prevention}, vol.~39,
  no.~2, pp. 347--355, mar 2007.

\bibitem{olaverri2016autonomous}
C.~Olaverri-Monreal, ``{Autonomous vehicles and smart mobility related
  technologies},'' \emph{Infocommunications Journal}, vol.~8, no.~2, pp.
  17--24, 2016.

\bibitem{8500367}
C.~{Olaverri-Monreal}, S.~{Kumar}, and A.~{Dìaz-Àlvarez}, ``Automated
  driving: Interactive automation control system to enhance situational
  awareness in conditional automation,'' in \emph{2018 IEEE Intelligent
  Vehicles Symposium (IV)}, 2018, pp. 1698--1703.

\bibitem{8317925}
J.~{Çapalar} and C.~{Olaverri-Monreal}, ``Hypovigilance in limited
  self-driving automation: Peripheral visual stimulus for a balanced level of
  automation and cognitive workload,'' in \emph{2017 IEEE 20th International
  Conference on Intelligent Transportation Systems (ITSC)}, 2017, pp. 27--31.

\bibitem{allamehzadeh2017cost}
A.~Allamehzadeh, J.~U. de~la Parra, A.~Hussein, F.~Garcia, and
  C.~Olaverri-Monreal, ``{Cost-efficient driver state and road conditions
  monitoring system for conditional automation},'' in \emph{IEEE Intelligent
  Vehicles Symposium (IV2017)}.\hskip 1em plus 0.5em minus 0.4em\relax IEEE,
  2017, pp. 1497--1502.

\bibitem{Bylykbashi2020}
K.~Bylykbashi, E.~Qafzezi, M.~Ikeda, K.~Matsuo, and L.~Barolli, ``{Fuzzy-based
  Driver Monitoring System (FDMS): Implementation of two intelligent FDMSs and
  a testbed for safe driving in VANETs},'' \emph{Future Generation Computer
  Systems}, 2020.

\bibitem{Goncalves2015}
J.~Goncalves, C.~Olaverri-Monreal, and K.~Bengler, ``{Driver Capability
  Monitoring in Highly Automated Driving: From State to Capability
  Monitoring},'' in \emph{IEEE Conference on Intelligent Transportation
  Systems, Proceedings, ITSC}, vol. 2015-Octob.\hskip 1em plus 0.5em minus
  0.4em\relax Institute of Electrical and Electronics Engineers Inc., oct 2015,
  pp. 2329--2334.

\bibitem{Takada2019}
S.~Takada, K.~Ueda, N.~Yamada, Y.~Iguchi, Y.~Tao, K.~Koizumi, and M.~Nakao,
  ``{Evaluation of Driver's Cognitive Load when Presented Information on the
  Windshield using P300 Latency in Eye-fixation Related Potentials},'' in
  \emph{2019 IEEE Intelligent Transportation Systems Conference, ITSC
  2019}.\hskip 1em plus 0.5em minus 0.4em\relax Institute of Electrical and
  Electronics Engineers Inc., oct 2019, pp. 4006--4011.

\bibitem{Noh2019}
Y.~Noh, S.~Kim, and Y.~Yoon, ``{Evaluation on Diversity of Drivers' Cognitive
  Stress Response using EEG and ECG Signals during Real-Traffic Experiment with
  an Electric Vehicle *},'' in \emph{2019 IEEE Intelligent Transportation
  Systems Conference, ITSC 2019}.\hskip 1em plus 0.5em minus 0.4em\relax
  Institute of Electrical and Electronics Engineers Inc., oct 2019, pp.
  3987--3992.

\bibitem{Petersen2019}
L.~Petersen, L.~Robert, J.~Yang, and D.~Tilbury, ``{Situational Awareness,
  Driver's Trust in Automated Driving Systems and Secondary Task
  Performance},'' \emph{SSRN Electronic Journal}, mar 2019.

\bibitem{Beller2013}
J.~Beller, M.~Heesen, and M.~Vollrath, ``{Improving the driver-automation
  interaction: An approach using automation uncertainty},'' \emph{Human
  Factors}, vol.~55, no.~6, pp. 1130--1141, dec 2013.

\bibitem{Verberne2012}
F.~M. Verberne, J.~Ham, and C.~J. Midden, ``{Trust in smart systems: Sharing
  driving goals and giving information to increase trustworthiness and
  acceptability of smart systems in cars},'' \emph{Human Factors}, vol.~54,
  no.~5, pp. 799--810, oct 2012.

\bibitem{Xiong2012}
\BIBentryALTinterwordspacing
H.~Xiong, L.~N. Boyle, J.~Moeckli, B.~R. Dow, and T.~L. Brown, ``{Use Patterns
  Among Early Adopters of Adaptive Cruise Control},'' \emph{Human Factors: The
  Journal of the Human Factors and Ergonomics Society}, vol.~54, no.~5, pp.
  722--733, oct 2012. [Online]. Available:
  \url{http://journals.sagepub.com/doi/10.1177/0018720811434512}
\BIBentrySTDinterwordspacing

\end{thebibliography}
\end{document}